\documentclass []{elsart}

\usepackage{graphicx}

\usepackage{amssymb}

\begin{document}
\begin{frontmatter}


\title{Asymmetric magnetization reversal in exchange biased polycrystalline F/AF bilayers
}
\author{D. Spenato\corauthref{cor}}
\corauth[cor]{Corresponding author: David Spenato LMB/UBO/CNRS UMR
6135, 6 avenue Le Gorgeu, 29285 Brest, France - fax +33 298 017
395 } \ead{david.spenato@univ-brest.fr}
\author{S.P. Pogossian and H. Le Gall }
\address{Laboratoire de Magn\'etisme de Bretagne, UBO/CNRS/UMR 6135, 6 avenue Le Gorgeu, 29285 Brest,
France}




\begin{abstract}
This paper describes a model for magnetization reversal in
polycrystalline Ferromagnetic/Antiferromagnetic exchange biased
bilayers. We assume that the exchange energy can be expanded into
cosine power series. We show that it is possible to fit
experimental asymmetric shape of hysteresis loops in exchange
biased bilayer for any direction of the applied field. The
hysteresis asymmetry is discussed in terms of energy
considerations. An angle $\beta$ is introduced to quantify the
easy axis dispersion of AF grains.\
\end{abstract}

\end{frontmatter}




\section{Introduction}
Exchange bias is a phenomenon known for a long time \cite{meik}
and characterized by a displacement (of shift value $H_s$), often
called "bias field", of the hysteresis loop along the field axis
as a result of exchange coupling between ferromagnetic (F) and
antiferromagnetic (AF) layers. The original model proposed by
Meiklejohn and Bean \cite{meik} is based on a perfect
uncompensated monolayer of spin at the surface of the
antiferromagnetic layer. This model predicts a shift field $H_{s}$
that is two orders of magnitude larger than that observed
experimentally. Several recent theories give improved predictions
of the magnitude of $H_{s}$, but they interpret in a different way
the physical origin of the effect. For example, Mauri \textit{et
al.} \cite{mauri} have shown that realistic values for $H_{s}$ can
be obtained if one assumes a domain wall formation in the AF layer
during the reversal of the F magnetization. Using localized atomic
spins, Koon \cite{koon} suggested 90 degrees spin-flop coupling
between the AF and F layers and predicted a correct magnitude of
$H_{s}$. A different approach was undertaken by Malozemoff
\cite{maloz} who assumed that the interface roughness greatly
reduces the number of uncompensated spins at the AF surface giving
rise to a smaller bias field. Siles and McMichael have proposed a
model for polycrystalline F/AF bilayers \cite{stilesb} consisting
of a F layer interacting with independent AF grains using three
contributions for the energy of each grain, a direct coupling, a
spin-flop coupling and a partial domain wall in the AF grains.
They showed that the unidirectional anisotropy comes from stable
grains as the magnetization is rotated while the hysteretic
effects come from grains where the AF order is switchable. In all
cases the spin-flop coupling does not contribute to the
unidirectional anisotropy. Despite many theoretical studies of the
origin and the magnitude of the bias field, a complete description
of the exchange anisotropy should incorporate explanations for the
other features of the exchange biased systems such as the
directional properties of the exchange bias and the strong
asymmetry of the hysteresis loop that occurs in these systems. The
angular dependence of $H_{s}$ and the coercivity $H_{c}$ was first
explored experimentally in NiFe/CoO bilayers \cite{ambrose}. The
variation of $H_{s}$ and $H_{c}$ versus the field angle
$\theta$$_{H}$ was described with cosine Fourier series, with odd
and even terms for $H_{s}$ and $H_{c}$ respectively instead of
simple sinusoidal functions as suggested initially \cite{meik}.
The angular dependence of the exchange anisotropy in F/AF bilayers
have been studied both theoretically and experimentally by several
authors. For some of them \cite{ambrose,cheng}, the complicated
angular behavior comes from the larger contribution of
higher-order terms $a_{n}$cos(n $\theta_{H}$)
 (n $\geq$ 3) in  $H_{s}$ and $H_{c}$. Kim \textit{et al.} \cite{kim} have modeled the
F/AF bilayer as a cubic structure and used an Heinsenberg
Hamiltonien to examine the angular dependence of perfect
interfaces. For compensated interface, their numerical results of
angular dependence of the exchange field were fit following the
work of Ambrose \textit{et al.} \cite{ambrose}, who have shown
that the angular dependence has a simple cosine function. For
rough interface the behavior is more complicated and simple cosine
form is suppressed and high order terms appear. Kim \textit{et
al.}\cite{kim} suggested that the possible origin of the higher
order terms may come from the nonrigid spin structure of the AF
layer. Geshev \textit{et al.} \cite{geshev} assumed a planar
domain wall formation at the antiferromagnetic side of the
interface with the reversal of ferromagnetic orientation
\cite{geshev}. Xi and White\cite{xiang} have shown that the
complex angular dependence of the exchange bias in NiFe/CrMnPt
bilayers can be understood by a simple Stoner-Wohlfath model
involving only a uniaxial and a unidirectional exchange coupling.
Very recently Krivorotov \textit{et al.}\cite{kriv} have shown
that the exchange anisotropy, induced by the F/AF exchange
coupling, consists  of a biaxial component, which gives rise to
the enhanced coercivity. In the latter model, the unidirectional
component is responsible for the shift of the hysteresis loop and
the threefold symmetry component for the symmetry breaking in the
magnetization reversal process. Tang \textit{et al.} \cite{tang}
have recently shown that the exchange bias effect, described by a
unidirectional anisotropy, is also accompanied by induced uniaxial
and fourfold in-plane contribution in Fe/MnPd exchange biased
bilayers. Most of these authors obtain good agreement between
theoretical and experimental angular variation of $H_{s}$ and
$H_{c}$.

Many authors have only fit the angular dependence
\cite{ambrose,kim}. Xi \textit{et al.} \cite{xiang} have used the
fit parameters for the angular dependence of the coercivity and
the exchange field to calculate the hysteresis loops for different
directions of the applied field. For the easy axis hysteresis
loop, the coercivity is about two time larger than the measured
one and the shifted hysteresis loop is symmetric. Our approach is
different, first we fit the easy axis hysteresis loop and then, by
using the fit parameters, calculate the angular dependence of the
coercivity and the shift field and also the hysteresis loop for
any angle of the applied field.

As to the asymmetry of the hysteresis loops in exchange biased
bilayers, there is a large variety of asymmetric shapes. The
modeling of hysteresis loops and especially its asymmetry was
achieved only for some particular shapes
\cite{mauri,stamps,stilescoer}. However none of these above
methods describe satisfactorily the shape of the measured easy
axis hysteresis loop presented on Fig.\ref{figure2}(a). This is
why we propose a theoretical model based on the development of the
energy functional for the exchange field in cosine powers series.
We show that the experimental asymmetric hysteresis loop measured
along the easy axis can be fit with our theoretical model. In the
present work, we study the magnetization reversal process using
conventional (VSM) magnetometry in NiFe/MnNi bilayers. The
theoretical magnetization curves, calculated for different
orientation of the applied field, are in good agreement with the
experimental data. We show that the asymmetry in the magnetization
reversal may be understood in terms of energetic considerations.

\section{Sample preparation and analysis}
Substrate$\backslash$$Ni_{81}Fe_{19} (520$\AA
$)\backslash$$Mn_{x}Ni_{100-x} (800$$\AA$)  bilayers were grown on
Corning Glass substrate  by RF diode sputtering using a standard Z
550 Leybold equipment with a magnetic field of 24 kA/m Oe applied
during deposition to induce an uniaxial anisotropy. The background
pressure was lower than $4\times10^{-7}$ mbar. Ni chips were
homogeneously added to a four inches diameter Mn target in order
to get films in the Mn composition range 5-80 percent. The
chemical homogeneity was verified by Electron Probe Micro Analysis
(EPMA) on several points of the sample. The Mn composition
variation is about one percent on the entire sample. As the
as-deposited samples did not exhibit exchange bias, after
deposition, they were annealed in a magnetic field of 80 kA/m,
aligned with the easy axis of the film, at $300^o$C for 5 hours to
induce the exchange field. The crystallographic structure was
examined by X-ray diffractometry with Cu K$\alpha$ radiation. A
(111) texture for the $Ni_{81}Fe_{19}$ was favored. Annealed
$Mn_{x}Ni_{100-x}$ films deposited on the underlying
$Ni_{81}Fe_{19}$ film were found to have a (111) texture with a
fct structure in the composition range 40 $<$x$<$80  that exhibits
exchange bias \cite{spen1}. The magnetic properties, such as the
saturation magnetization  $M_{s}$, were obtained from
magnetization loops (M-H loops) measured at room temperature using
a VSM. It is well established that the shift of the hysteresis
loop of exchange biased bilayers depend on intrinsic properties
such as the AF layer thickness, the F layer thickness and the
deposition conditions (for a review see \cite{revex}). All these
parameters induce hysteresis loop with a more or less pronounced
asymmetry. For our study we have chosen samples with a
well-defined magnetization reversal asymmetry  observed by easy
axis hysteresis loop. In our samples the forward loop shows a
slope, as if the full magnetization reversal takes place slowly,
while the recoil loop is very square which may be the consequence
of either coherent magnetization flip (such as in monodomain
particles) or the presence of high mobility domain walls.

\section{Model}
The magnetization reversal that occurs in exchange biased bilayers
can not be explained with a simple algebraic addition of an
exchange field to the applied magnetic field $H_{a}$ because of
the asymmetric shape of the hysteresis loop. In order to take into
account the exchange bias, Meiklejohn and Bean \cite{meik} have
proposed a linear term that accounts for the hysteresis curve
shift, but which does not explain the asymmetry of the hysteresis
loops for exchange biased F/AF systems. Owing to the large variety
of hysteresis asymmetric curves, one should admit that one
parameter exchange energy is no longer sufficient to interpret the
form of these curves. In order to take into account the complex
physical phenomena giving rise to the asymmetry of hysteresis
curve, other terms should be added to the exchange energy
expression describing the F/AF exchanged biased samples. Several
authors added a non linear biquadratic term proposed by
Slonczewski\cite{slon}. However, that term is not sufficient to
account for a large variety of asymmetric hysteresis curves known
for various systems of F/AF samples. Based on symmetry
considerations Ambrose \textit{et al.}\cite{ambrose} proposed a
Fourier odd and even series development of the shift field and the
coercivity respectively that accounts for the angular variation of
$H_{s}$ and $H_{c}$. More recently, using the Stoner-Wohlfath
model, involving only a uniaxial and a unidirectionnal exchange
coupling, Xi \textit{et al.} \cite{xiang} fitted the hard axis
hysteresis loop to experimental data. Nevertheless, the easy axis
hysteresis loop shape fitted poorly with experimental curve.
Different mechanisms of magnetization reversal such as, the AF
domain wall formation, the different direction of easy axis in AF
and F layers, the grain formation etc contribute to the hysteresis
curve shape. Due to the complexity of magnetization reversal
processes, which are strongly influenced by the physical state of
F/AF interface and therefore very sensitive to the physical
processes of sample preparation, a large number of Fourier series
terms should be taken into account in order to achieve an
agreement between the experimental data and theoretical
calculations. In general, Fourier series convergence is rather
slow. This is why we propose a cosine power series expansion of
the exchange energy. Moreover, the grain formation and the easy
axis distribution of different grains with respect to the
magnetization direction of F layer leads us to introduce a mean
easy axis direction of different AF grains. The exchange coupling
between different grains, and the way they are influenced by F
layer, suggests that statistical mean easy axis of AF grains may
deviate from that of F layer. The exchange coupling with the AF
layer is assumed to result in a single domain in the F film as
recently observed in $NiFe/Fe_{50}Mn_{50}$ by magneto-optic Kerr
effect \cite{zhou}. Therefore, we admit a uniform switching of the
total magnetization of the F layer. The magnetic configuration of
the F/AF bilayer is shown in Fig.\ref{figure1}. The magnetic
energy per unit area of exchange coupled F layer
 and AF layers can be written as follows:

 \begin{equation} \label{en}
 E =
 K_{F}t_{F}sin^{2}\varphi-H_{a}M_{F}t_{F}cos(\theta-\varphi)-E_{e}(\varphi-\beta)
 \end{equation}

 where $K_{F}$ is the anisotropy constant of the F layer with its
 magnetization $M_{F}$ making an angle $\varphi$ with respect to the
 the F easy axis. The first term represents the anisotropy energy
 of the F layer. The second term corresponds to the Zeeman energy
 of the F layer submitted to an external static field $H_{a}$
 applied at an angle $\theta$ relative to the anisotropy easy axis
 of the F layer. The last term represents the exchange anisotropy energy,
 where $\beta$ is the angle of the AF mean statistical easy axis with respect to the F easy
 axis.
 $E_{e}$($\varphi$-$\beta$) may be written as:

\begin{equation}\label{funct}
 E_{e}(\varphi-\beta)=
\sum_{n}J_{n}cos^{n}(\varphi-\beta)
 \end{equation}

If one admits, \emph{a priori}, that $E_{e}(\varphi-\beta)$ is
linear: $E_{e}(\varphi-\beta)$=$J_{1}cos(\varphi-\beta)$ then that
represents the simple unidirectional energy originally proposed by
Meiklejohn and Bean \cite{meik} (with $\beta$ = 0) and used
recently by Xi \textit{et al} \cite{xiang} (with $\beta$ $\neq$
0). The biquadratic term of  $E_{e}(\varphi-\beta)$ was already
used by Slonczewski \cite{slon} and later by Stamps \cite{stamps}
to describe the coupling of the F layer to both sublattices of two
sublattice antiferromagnet. Sometimes, based on crystal symmetry
considerations, threefold and fourfold components were used in
$E_{e}(\varphi-\beta)$ energy functional. The threefold component
was recently found to be responsible for the symmetry breaking in
the magnetization reversal process in $Fe/MnF_{2}$\cite{kriv}.
Tang \textit{et al.} \cite{tang} have shown that the exchange bias
effect, described by a unidirectional anisotropy, is also
accompanied by induced uniaxial and fourfold in-plane contribution
in Fe/MnPd exchange biased bilayers. Because of the complexity of
the exchange bias phenomenon, we can not limit our development to
any finite number of terms. So the total energy in terms of
effective field can be written as:

 \begin{equation} \label{red}
 \frac{E}{t_{F}M_{F}} =
 \frac{H_{K}}{2}sin^{2}\varphi-H_{a}cos(\theta-\varphi)-\sum_{n}H_{n}cos^{n}(\varphi-\beta).
 \end{equation}

Where $H_{K}$ is the anisotropy field and the $H_{n}$ are the
coefficients of the power series development. Based on the above
model, the magnetization curves of the F/AF bilayer are obtained
by numerical calculations.

The coefficients H$_1$, H$_2$, H$_3$ etc... being the power
expansion coefficients of energy functional, do not reflect the
overall energy symmetry. We have truncated cosine power series at
n= 5 for which the fit to the shape of easy and hard axis
hysteresis curves was the best. The fitting to the easy axis
hysteresis loop improves when one takes into account more terms.
However, in that case, the hysteresis loop calculated for
$\theta$$\neq$0 by using fit parameters of easy axis hysteresis
loop shows parasitic minor loops. One may obtain better fit by
using 10 or 20 series terms. But that will not mean anyhow that
the crystal has a 10 th or 20-th order symmetry since only the sum
of all terms should provide such or such symmetry and not the each
term. When we truncated the expansion to n = 4 the best fit to
easy axis hysteresis curve becomes square.

To give a physical signification to the 5 th order coefficient is
difficult since our samples were prepared with RF sputtering and
none of the textures were pronounced. After a thermal annealing,
an interdiffusion have been induced \cite{chen}. Moreover, as
reported in a previous work \cite{spen1}, a ternary NiMnFe alloy
seems to be formed in the F/AF interface region. Chen \emph{et
al.}\cite{chen} showed that the interdiffusion alters dramatically
the exchange bias and that the hysteresis loop shape becomes
asymmetrical. To the best of our knowledge there is no systematic
study concerning on the influence of the alloy formed at the
interdifusion layer at the F/AF interface on the exchange energy
functional. So it is a difficult task to associate a physical
meaning with the exchange energy expansion coefficients.

\section{Results and discussions}
In a first step we have fitted the hysteresis loop to the
theoretical model given by Eq. \ref{red}, when the field is
applied along the easy axis. The comparison between theoretical
and experimental loops are presented on Fig. \ref{figure2}(a). One
can observe a strong asymmetry between the left and right side of
the measured hysteresis loop. The magnetization reversal on the
left side of the loop exhibits a hard-axis-type slope while the
reversal for the right-hand side of the loop occurs with a
vertical branch of the loop. It can be observed that the
theoretical magnetization curve fits  well the measured hysteresis
loop. The model gives values expected either for the hysteresis
loop shift ($H_{s}$ = 950 A/m) and coercivity ($H_{c}$ = 900 A/m)
or for the strong asymmetry in the magnetization reversal. The
values of the fit parameters are $\beta$=-0,064, $H_{K}$ = 533
A/m, $H_{1}$ = 1661 A/m, $H_{2}$ =-903 A/m, $H_{3}$ =616 A/m,
$H_{4}$ =501 A/m, $H_{5}$ =-599 A/m. The first term $H_{K}$ is
about two times larger than the anisotropy field of the
as-deposited NiFe layer (i.e. 360 A/m). The second term $H_{1}$,
that may be related to the exchange biased field, is about 1.5
larger than the hysteresis loop shift.

One can observe a distinct kink on the left side of the calculated
magnetization loop whereas the experimental data show a smooth
variation. Leigthon \textit{et al.}\cite{kink} attributed that
kind of kink to a two-stages magnetization reversal. The defect
concentration is of high importance for exchange bias
characterization as indicated by several authors (see the review
by Stamps \cite{stamps}). The higher order Fourrier coefficients
in the exchange energy functional of  single crystal samples are
generally much larger than those for the polycrystalline samples.
In polycrystalline materials the situation is complicated due to
the spread in AF grain concentration. It is further complicated in
the presence of an interdiffusion layer that may be a ternary
NiMnFe alloy. According to Mewes \emph{et al.} \cite{mewes}  the
Stoner-Wolfarth model shows sharp edges which are rounded in the
experiment for polycrystalline samples. This may explain the
smooth variation of our experimental results.

Energy diagrams are often useful to explain the magnetization
reversal in magnetic hetorostructures. Fig. \ref{figure3}
represents the left (Fig. \ref{figure3}(a)) and the right (Fig.
\ref{figure3}(b)) side of the measured hysteresis loop. In our
calculations, for each value of the applied field, the system is
in a local energy minimum which is plotted for the forward (left
side) and reverse (right side) loops on the same figure. The
energy curve presents an hysteretic behavior as a function of the
applied field. On the same figure one can observe a strong
asymmetry in the magnetization reversal due to the slow variation
of energy local minimum as a function of applied field. On the
right side of the figure (Fig. \ref{figure3}(b)), there is an
abrupt lowering of local energy minimum at about $H_{a}$=0 which
corresponds to an abrupt change of the energy minimum angle
($\varphi$). At the left side of the loop, the magnetization must
overcome an energy barrier that vanishes after an increase
$\Delta$H of the field (Fig. \ref{figure3}(a)). This reversal may
occur with a number of discontinuous jumps giving rise to "kinks"
in the hysteresis loop. For the reverse loop, the magnetization
vector flips rapidly from $\varphi$=$\pi$ to $\varphi$=0 at a
critical field about $H_{a}$$\simeq$0.

We have used the same fit parameters to calculate the
magnetization curves for an applied field making different angles
($\pi$/3, and $\pi$/2) with the F easy axis. The results are
presented in Fig. \ref{figure2}(b and c). As observed on the Fig.
2( b and c), the small deviation of experimental and theoretical
curves is attributed to the initial dispersion of the as-deposited
NiFe anisotropy revealed by a small coercivity measured along the
hard axis. That may come from the aligning field in the sputtering
chamber which lacks uniformity and may give rise to the F
anisotropy dispersion. Indeed we did not include in our model such
a dispersion term which is the main cause of that difference. Fig.
\ref{figure4} describes the comparison of theoretical and
experimental angular variation of the coercivity $H_{c}$ Fig.
\ref{figure4}(a) and the hysteresis loop shifts $H_{s}$ Fig.
\ref{figure4}(b). The fit parameters correspond to the easy axis
hysteresis loop. The behavior of the coercivity is very similar to
that found by Ambrose \textit{et al.}\cite{ambrose}. The measured
coercivity does not vanish in the directions different from that
of the easy axis ($\theta$$\neq$0). For some angles there is
discrepancies between the computed hysteresis shift and the
measured one. First, the model predicts maxima of $H_{s}$ located
at about $\theta$ = $\pm$$\pi$/15 and $\pi$$\pm$2$\pi$/3. This
behavior is different from the believed one, according to which
the largest values of $H_{s}$ are expected at $\theta$ = 0 and
$\pi$ as mentioned by several authors
\cite{ambrose,xiang,stilescoer}. We observed experimentally the
largest values of  $H_{s}$ for $\theta$ = 0 and $\pi$. Our angular
dependence of the coercivity and hysteresis loop shift differs
from the published results \cite{ambrose}. In our calculations the
theoretical angular variation of $H_{c}$ and $H_{s}$ is obtained
from a unique set of parameters deduced by the fit to the easy
axis hysteresis loop instead of two different fits for the
coercivity and the exchange field as usually presented
\cite{ambrose,cheng}.

We think that, the discrepancy between theoretical and
experimental results in Fig.\ref{figure4} is due to the fact that
our samples are polycrystalline. Riedling \emph{et al}
\cite{riedling} show that a complex H$_S$($\theta$) behaviour may
be due to the epitaxially grown samples. Liu \emph{et al}
\cite{liu} realised epitaxially grown NiFe/FeMn exchange biased
bilayers on Si(100) films buffered by Cu. LEED experiments
supported the existence of a six-fold symmetry. The obtained
H$_S$($\theta$) and H$_c$($\theta$) curves were strongly dependent
on the growth technique. Polycrystalline samples show perfect
cos($\theta$) behaviour even in the presence of a well pronounced
induced (111) texture. That work supports our experimental data
where also a well defined cos($\theta$) behaviour was observed.
The epitaxial samples in the work of liu \emph{et al}. show that
H$_S$($\theta$) and H$_c$($\theta$) curves have several minima for
different Cu buffer layer.  The behaviour of H$_S$($\theta$) and
H$_c$($\theta$) curves shapes may also depend on the existence of
an interdiffusion layer, but we have not sufficient experimental
data to confirm or exclude such an influence.

An important feature of the present model is the non zero value of
$\beta$=-0.064 which is the angle between the F and AF easy axis
directions. We suppose that it is due to the AF grains formation
and the local exchange energy variation at the interface of F/AF
layers. Owing to the interface imperfections, the easy axis of AF
grains will suffer some dispersion \cite{fulcom,stilesb}. A
statistical mean easy axis direction can be a useful parameter to
quantify that dispersion. The physics of such dispersion is
somewhat complicated due to the lack of information about, the
exchange energy local variation, the roughness characteristics and
its influence on the energetic balance of concurrent energetic
processes for energy minimization. This is why the value of
$\beta$, \textit{a priori}, may be slightly different from zero.
So we think that it can be an additional pertinent parameter to
describe the exchange bias.

\section{Conclusions}
In summary, we have presented a model based on the representation
of the exchange energy by cosine power series. We have shown that
it is possible to fit the asymmetric shape of the easy axis
hysteresis loop. The asymmetric hysteresis loop shape is
discussed on the basis of energy considerations. An angle $\beta$
is introduced to quantify the easy axis dispersion of AF grains.
The easy axis fit parameters describe satisfactorily the
hysteresis curves in other directions.

\section*{Acknowledgments}
The authors thank J. Ben Youssef for samples preparation.

 \clearpage

\begin{figure}
\begin{center}
\caption{Vector diagram for an exchange coupled F/AF bilayer
submitted to an applied field $H_{a}$} \label{figure1}
\end{center}
\end{figure}

\begin{figure}
\begin{center}
\caption{Representative hysteresis loops for a
$Ni_{81}Fe_{19}$(520$\AA$)/$Mn_{46}Ni_{54}$(800$\AA$)bilayer at
(a) 0, (b) $\pi$/3, (c) $\pi$/2 of the applied field referred to
the easy axis. The experimental data are shown by scattered empty
dots. The solid lines are the theoretical magnetization curves. }
\label{figure2}
\end{center}
\end{figure}

\begin{figure}
\begin{center}
\caption{(a): Forward measured magnetization hysteresis loop
(upper) and corresponding calculated
 energy hysteresis loop (lower). (b): reverse measured magnetization hysteresis loop
(upper) and corresponding calculated
 energy hysteresis loop (lower). } \label{figure3}
\end{center}
\end{figure}

\begin{figure}
\begin{center}
\caption{Angular dependence of (a) shift field $H_{s}$ and (b)
coercivity of an exchange coupled NiFe/MnNi bilayer. The open dots
are the experiments and the solid line is the theory.}
\label{figure4}
\end{center}
\end{figure}

\clearpage

\begin{figure}
\begin{center}
\includegraphics[width=10 cm]{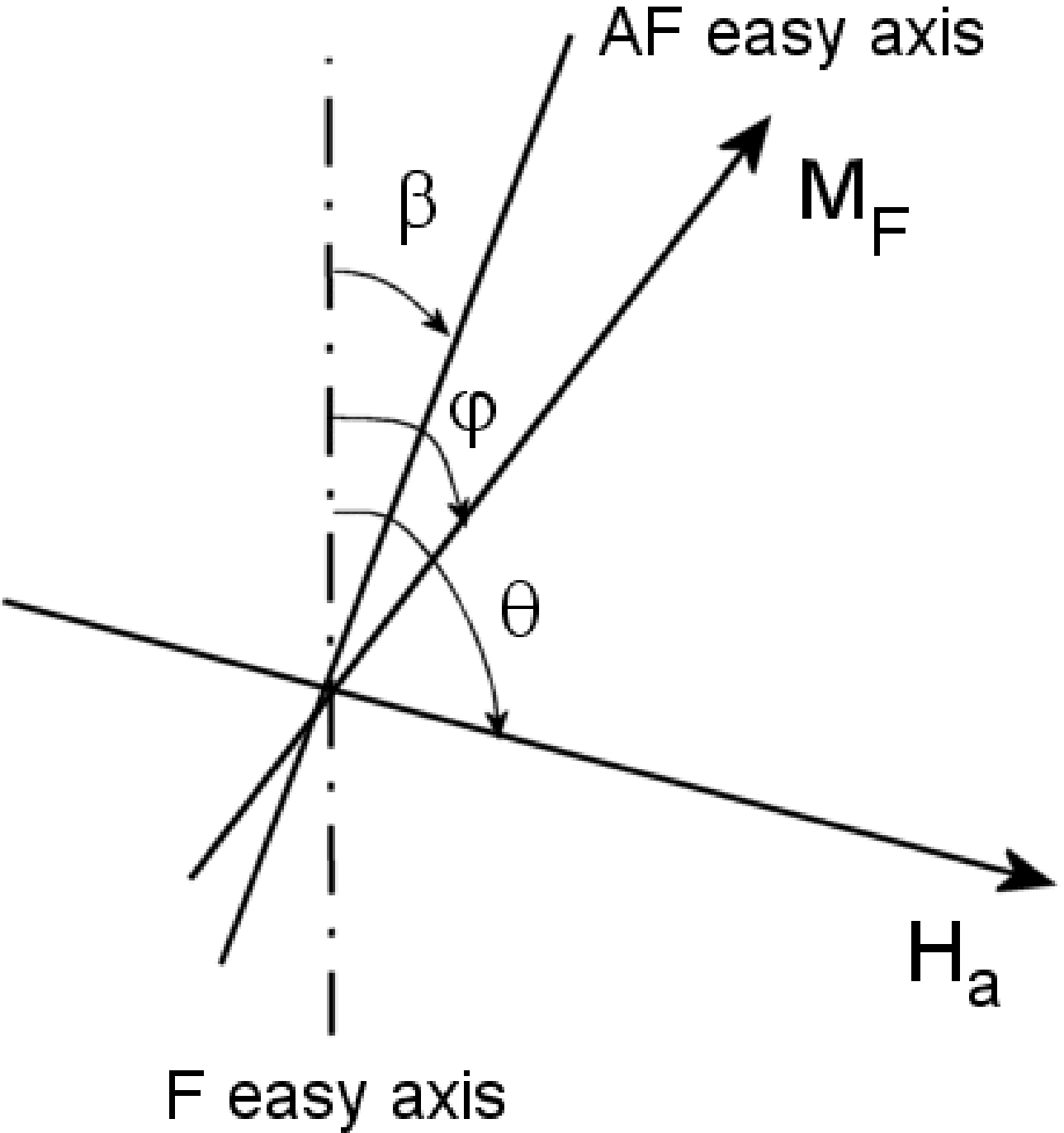}

\end{center}

\end{figure}
\hspace{6.5cm} Figure \ref{figure1} \\ \\ \\ \\ \\ \\ \\ \\   D.
Spenato, S. P. Pogossian and H. Le Gall

\clearpage

\begin{figure}
\begin{center}
\includegraphics[width=13 cm]{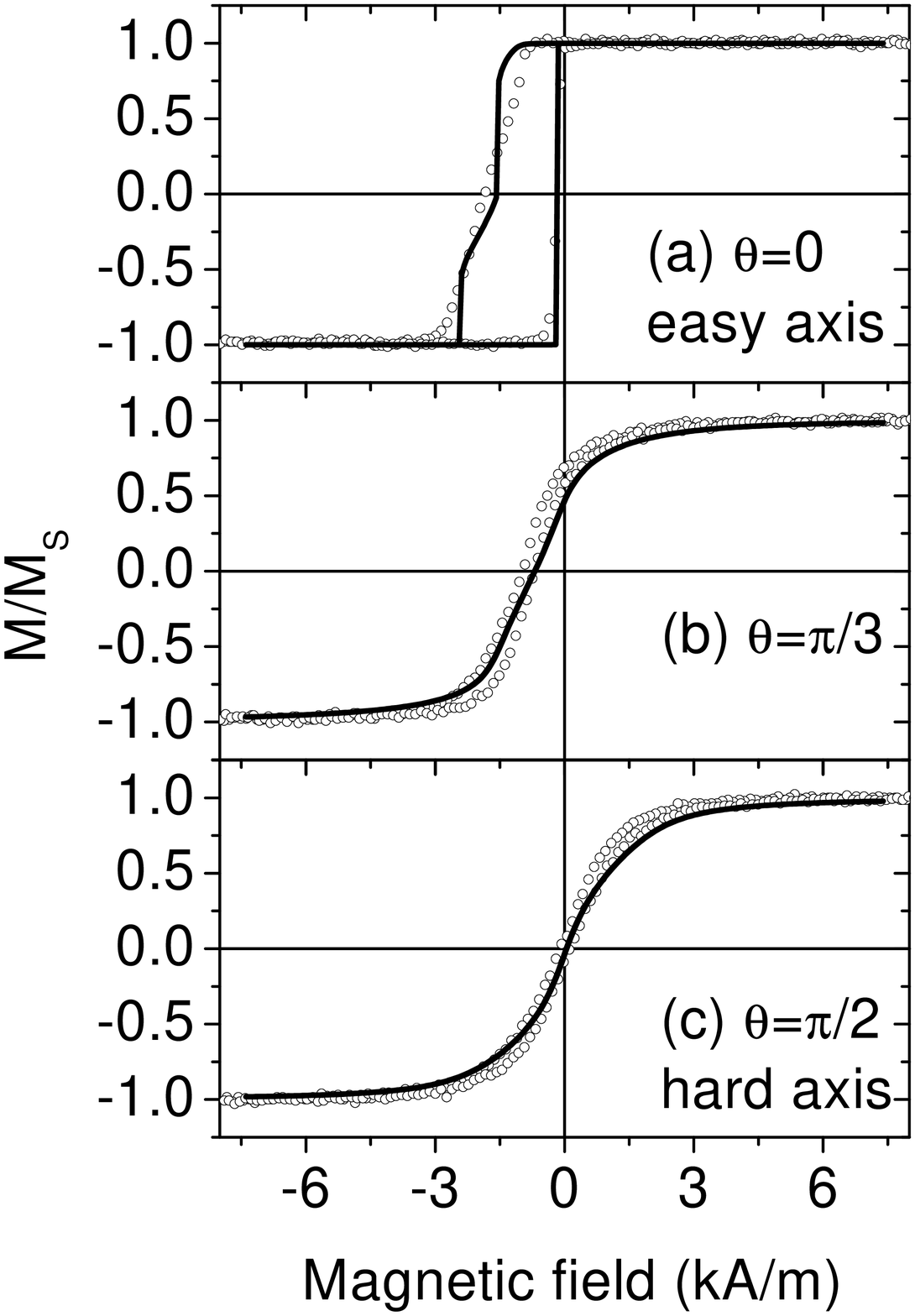}
\end{center}
\end{figure}
\hspace{6.5cm} Figure \ref{figure2}\\ \\ \\
 D. Spenato, S. P. Pogossian and H. Le Gall

\clearpage

\begin{figure}
\begin{center}
\includegraphics[width=15 cm]{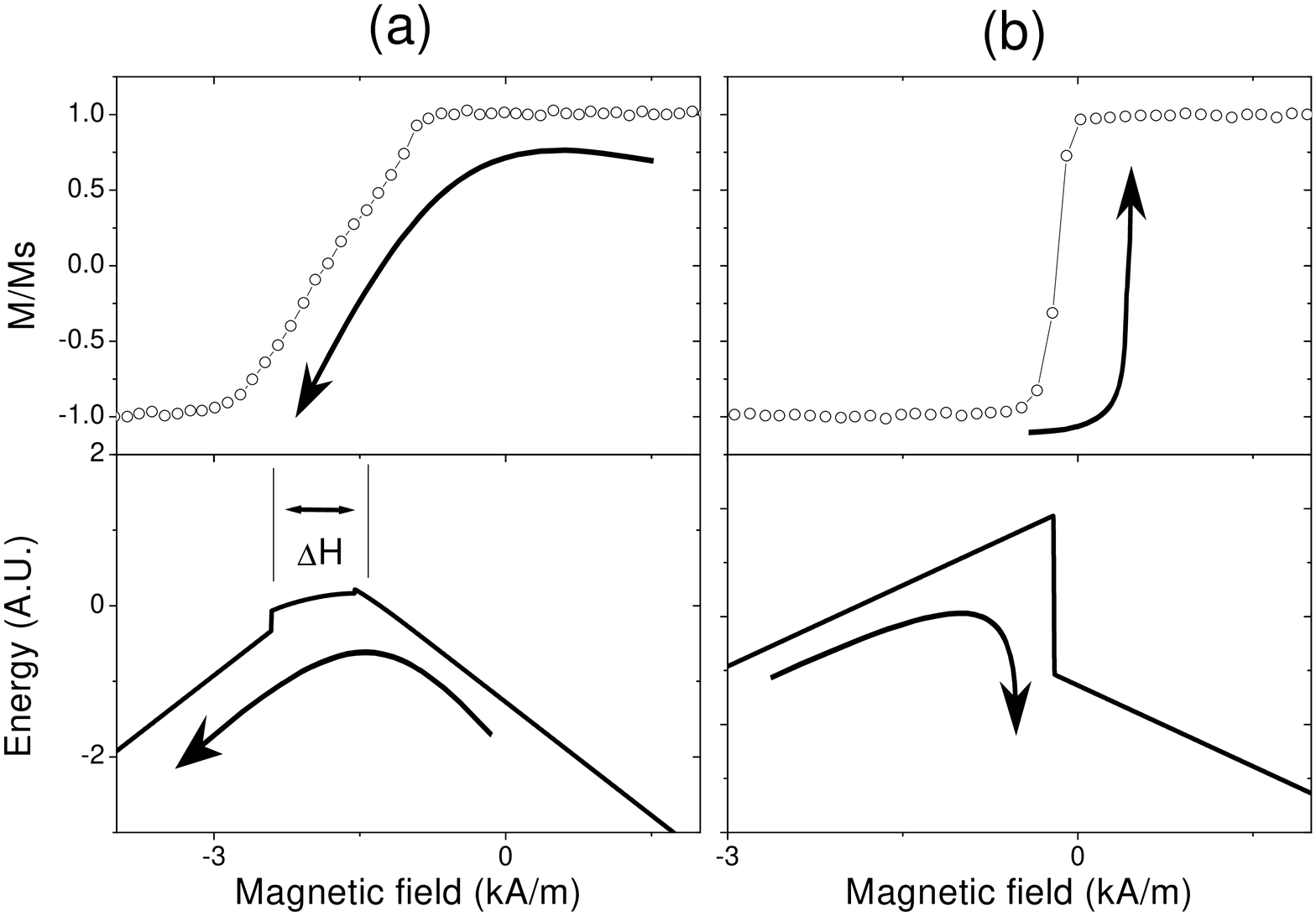}
\end{center}
\end{figure}
\hspace{6.5cm} Figure \ref{figure3}\\ \\ \\ \\ \\ \\ \\ \\
 D. Spenato, S. P. Pogossian and H. Le Gall

\clearpage

\begin{figure}
\begin{center}
\includegraphics[width=13 cm]{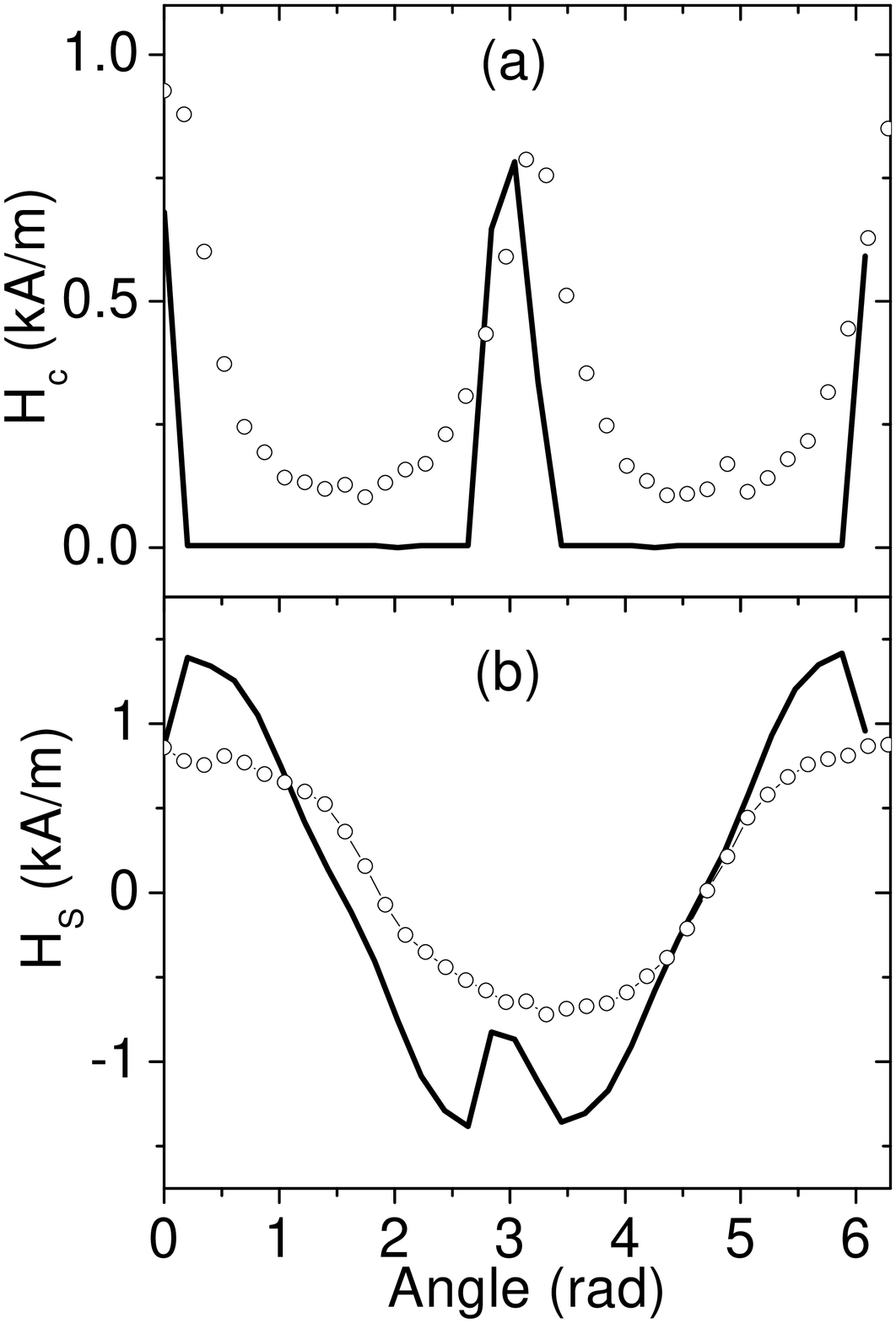}
\end{center}
\end{figure}
\hspace{6.5cm} Figure \ref{figure4}\\ \\ \\
  D. Spenato, S.P.
Pogossian and H. Le Gall

\bibliographystyle{elsart-num}

\bibliography{referenc}


%
%

%
%

\end{document}